\documentclass[fleqn,usenatbib]{mnras}

\usepackage{mathptmx}
\usepackage{txfonts}
\usepackage{float}
\usepackage[T1]{fontenc}

\DeclareRobustCommand{\VAN}[3]{#2}
\let\VANthebibliography\thebibliography
\def\thebibliography{\DeclareRobustCommand{\VAN}[3]{##3}\VANthebibliography}

\usepackage{graphicx}
\usepackage{amssymb}	

\title[Updated values of solar gravitational moments]{Updated values of solar gravitational moments $J_{2n}$ using HMI helioseismic inference of internal rotation}

\author[R. Mecheri et al.]{
R. Mecheri,$^{1}$\thanks{E-mail: redouane.mecheri@craag.edu.dz (KTS)}
M. Meftah$^{2}$
\\
$^{1}$Centre de Recherche en Astronomie, Astrophysique et G\'{e}ophysique,
CRAAG, BP 63, 16340 Bouzar\'{e}ah, Algiers, Algeria\\
$^{2}$Laboratoire Atmosph\`{e}res, Milieux, Observations
Spatiales (CNRS-LATMOS), 11 Boulevard d'Alembert, 78280, Guyancourt, France
}

\date{Accepted XXX. Received YYY; in original form ZZZ}
\pubyear{2015}

\begin{document}
\label{firstpage}
\pagerange{\pageref{firstpage}--\pageref{lastpage}}
\maketitle

\begin{abstract}

The solar gravitational moments $J_{2n}$ are important astronomical quantities whose precise determination is relevant for solar physics, gravitational theory and high precision astrometry and celestial mechanics. Accordingly, we propose in the present work to calculate new values of $J_{2n}$ (for $n$=1,2,3,4 and 5) using recent two-dimensional rotation rates inferred from the high resolution SDO/HMI helioseismic data spanning the whole solar activity cycle 24. To this aim, a general integral equation relating $J_{2n}$ to the solar internal density and rotation is derived from the structure equations governing the equilibrium of slowly rotating stars. For comparison purpose, the calculations are also performed using rotation rates obtained from a recently improved analysis of SoHO/MDI heliseismic data for solar cycle 23. In agreement with earlier findings, the results confirmed the sensitivity of high order moments ($n>1$) to the radial and latitudinal distribution of rotation in the convective zone. The computed value of the quadrupole moment $J_{2}$ ($n=1$) is in accordance with recent measurements of the precession of Mercury's perihelion deduced from high precision ranging data of the MESSENGER spacecraft. The theoretical estimate of the related solar oblateness $\Delta_{\odot}$ is consistent with the most accurate space-based determinations, particularly the one from RHESSI/SAS.

\end{abstract}


\section{Introduction}\label{sec:introduction}

Solar gravitational moments $J_{2n}$ are coefficients that describe the rotation-induced deviation of the Sun's outer gravitational potential $\phi_{out}$ from a spherical configuration. Assuming an axial symmetry around the rotation axis, they intervene in the expression of $\phi_{out}$ as projection coefficients on the basis of Legendre polynomials:
\begin{equation}\label{eq:phiout}
  \phi_{out}(r,u)=-\frac{GM_{\odot}}{r}\left[1-\sum_{n=1}^{\infty}
  \left(\frac{R_{\odot}}{r}\right)^{2n}J_{2n}P_{2n}(u)\right]
\end{equation}
The odd terms have been omitted from the series in equation~(\ref{eq:phiout}) because of equatorial symmetry. The quantities $G$, $M_{\odot}$, $r$, $R_{\odot}$, $P_{2n}$ and $u$=$cos\theta$, are respectively the gravitational constant, the solar mass, the distance from the centre of the Sun, the mean solar radius, the Legendre polynomials of degree 2$n$ and the cosine of the colatitude of the Sun $\theta$ (angle to the rotation axis). The accurate determination of $J_{2n}$ is of interest not only in solar physics but also in many other astrophysical applications. The most famous one is undoubtedly the test of general relativity (GR) resulting from the combination of the value of the quadrupole moment $J_{2}$ with the measurements of the anomalous precession of Mercury's orbit \citep{dicke1964,shapiro1972,campbell1983,lydon1996,chapman2008,gough2013}. In the same way, $J_{2}$ can be used to constraint the Eddington-Robertson parameters in the Parametrized-Post-Newtonian (PPN) theory of gravity, an alternative gravitation theory to GR \citep{pireaux2003,iorio2005}. In astrometry, an estimate of $J_{2}$ makes possible to study its effect on the astrometric \citep{kislik1983,bursa1986} and celestial mechanics \citep{YanXu2011,YanXu2017,vaishwar2018} determination of planetary orbits and also on the dynamics of the earth-moon system \citep{bois1999}. For detailed reviews on the implication of $J_{2n}$ in alternative theories of gravitation, high precision astrometry and celestial mechanics, readers are referred to the two articles by \citet{rozelot2009,rozelot2013}. In solar physics, $J_{2n}$ indicate non-uniform mass and angular velocity distribution inside the Sun and their accurate knowledge would provide a good constraint on internal structure and rotation \citep{dicke1967,ulrich1981a,ulrich1981b,paterno1996,godier1999,armstrong1999,mecheri2004}, and on solar cycle models through the study of their temporal evolution \citep{antia2008}, complementing thus the constraints imposed by helioseismology.

Several observational and theoretical works have been undertaken to determine solar gravitational moments $J_{2n}$ (mainly $J_{2}$). In general, the observational determinations are either from oblateness estimates based on the profile of the Sun's limb (\citet{dicke1967,dicke1986} using the Solar Distortion telescope, \citet{hill1975} using the SCLERA telescope, \citet{lydon1996} using the Solar Disk Sextant (SDS) instrument, \citet{rosch1996,rozelot1997} using the Pic du Midi heliometer, \citet{fivian2008} using the Solar Aspect Sensor (SAS) onboard of the Reuven Ramathy High-Energy Solar Spectroscopic Imager (RHESSI) satellite), or from astrometric observations of planetary orbit of Mercury and other minor planets such as Icarus \citep{lieske1969,anderson1978,afanaseva1990,landgraf1992, pitjeva2005} or form Lunar Laser Ranging (LLR) data \citep{rozelot1998}. Theoretical expressions relating the solar gravitational moments $J_{2n}$ to the inner structure and dynamics of a star can be determined using the theory of slowly rotating stars \citep{schwarzschild1947,sweet1950}. Early application of this theory to the Sun was done by \citet{roxburgh1964,goldreich1967,gough1981} in the context of analyzing internal rotation. It was used for the determination of $J_{2n}$ by \citet{ulrich1981a,ulrich1981b} using a simple quadratic rotation law. Several theoretical determinations followed \citeauthor{ulrich1981a} work, using two-dimensional helioseismically inferred rotation rates either in a parametric form \citep{paterno1996,godier1999,roxburgh2001,mecheri2004} or through direct inversion of rotational frequency splitting \citep{gough1982,campbell1983,duvall1984,brown1989,pijpers1998,armstrong1999,antia2000,antia2008}.
All these contributions computed values of $J_{2n}$ either from a differential or an integral equation which was derived explicitly for the special case of $n$=1 or $n$=2. Exception is made to works by \citet{armstrong1999,roxburgh2001} and particularly \citet{mecheri2004} who derived a convenient general form of the Poisson equation whose solution at the surface gives $J_{2n}$ for any value of $n$.

In the present work, we take over the above mentioned equation \citep[see][equation~(4)]{mecheri2004} and perform further algebraic calculations to derived a general integral equation relating $J_{2n}$ to the internal rotation following the Green's functions method described by \citet{pijpers1998}. This integral equation is then used to compute values of $J_{2n}$ for $n$=1, 2, 3, 4, and 5 taking into account new constraints on internal rotation provided by the high resolution HMI (Helioseismic and Magnetic Imager) aboard of SDO (Solar Dynamics Observatory) helioseismic data covering the whole solar cycle 24. Our main equations are presented in Section~\ref{sec:model}. The results of our computations of $J_{2n}$ are presented and discussed in Section~\ref{sec:results}. Finally, we give our principal conclusions in Section~\ref{sec:conclusions}.

\section{General integral equation for $J_{2\lowercase{n}}$}
\label{sec:model}
\begin{figure}
	\includegraphics[height=5.5cm]{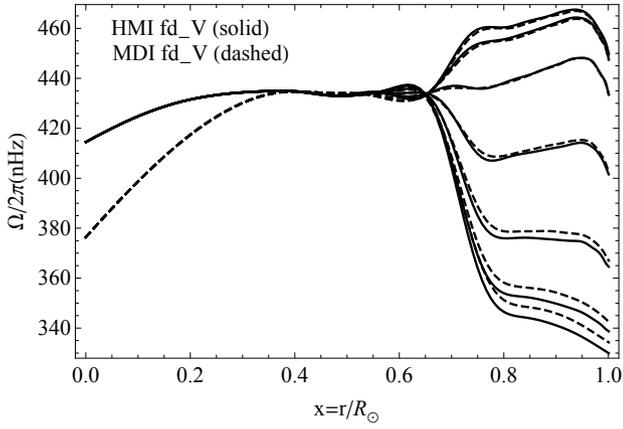}
    \caption{Time-averaged radial profiles of HMI (solid lines) and MDI (dashed lines) rotation, obtained from helioseismic data of full disk (fd\_V) dopplergrams, given each $15^{\circ}$ from equator (top) to pole (bottom).}
    \label{fig:figrot}
\end{figure}
\begin{figure*}
  \includegraphics[width=17.7cm]{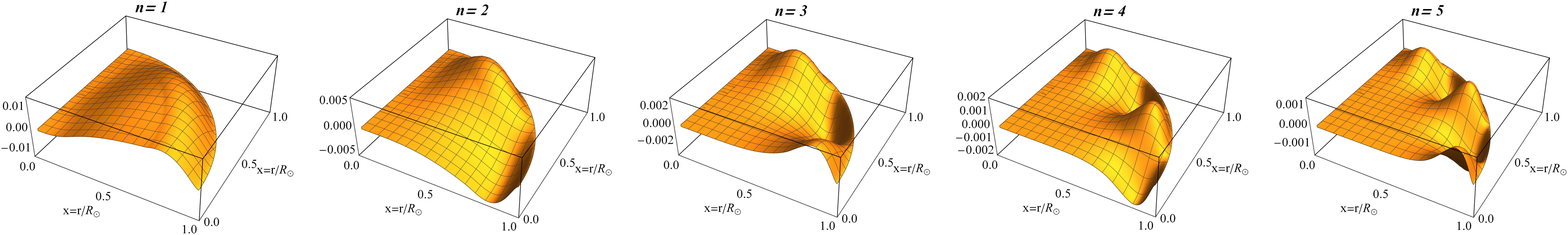}
  \includegraphics[width=17.7cm]{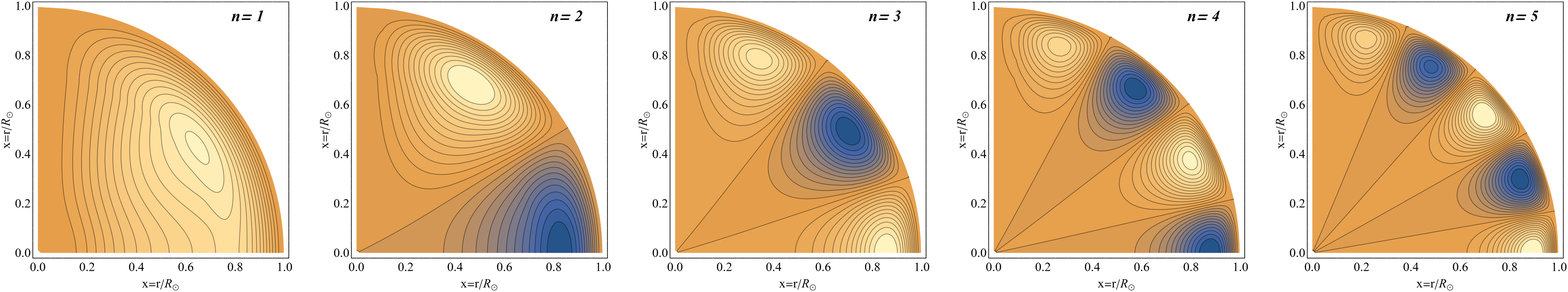}
  \caption{Three-dimensional plots (top panels) of the normalized kernel $F_{2n}$ as a function of $x=r/R_{\odot}$ and latitude for $n$=1,2,3,4 and 5 and their corresponding contour plots (bottom panels).}
    \label{fig:figF2n}
\end{figure*}
\begin{figure*}
  \includegraphics[width=17.7cm]{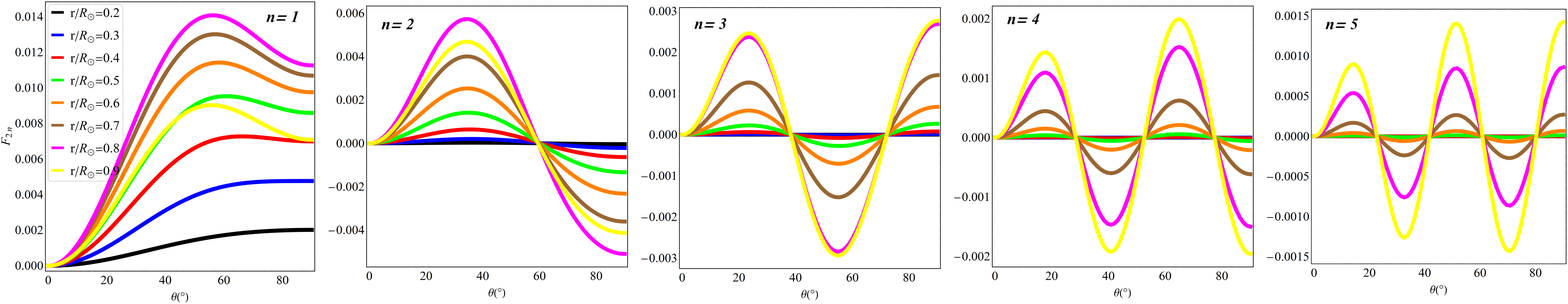}
  \includegraphics[width=17.7cm]{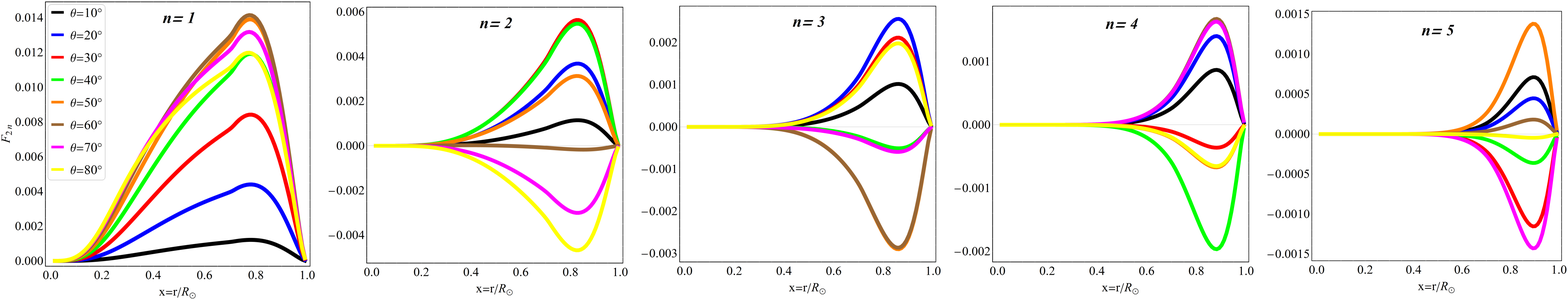}
  \caption{Plots of latitudinal (top panels) and radial (bottom panels) cuts of the normalized kernel $F_{2n}$ for $n$=1,2,3,4 and 5, respectively for different values of $x=r/R_{\odot}=0.2,~0.3,~0.4,~0.5,~0.6,~0.7,~0.8$ and $0.9$ and colatitude $\theta(^{\circ})=10^{\circ},~20^{\circ},~30^{\circ},~40^{\circ},~50^{\circ},~60^{\circ},~70^{\circ}$ and $80^{\circ}$.}
    \label{fig:figF2nrt}
\end{figure*}

Theoretical expressions relating the distortions of a star to the internal mass, density and rotation can be obtained under the assumption of a slow rotation (i.e. centrifugal acceleration small compared to the gravitational acceleration) where all stellar structure quantities are described in terms of perturbations (with subscript 1) of the spherically symmetric non-rotating star (with subscript 0). The perturbations are thereby expanded on the basis of Legendre polynomials giving a gravitational potential inside the Sun $\phi_{int}$ as follow:
\begin{equation}\label{eq:phiint}
  \phi_{int}(r,u)=\phi_{0}(r)+\phi_{1}(r,u)=\phi_{0}(r)+
  \sum_{n=1}^{\infty}\phi_{12n}(r)P_{2n}(u)
\end{equation}
where $\phi_{0}$ is the gravitational potential of a spherical Sun and $\phi_{12n}$ represent the projections of the perturbed gravitational potential $\phi_{1}$ on the Legendre polynomials basis. The gravitational moments $J_{2n}$ are given assuming the continuity of the gravitational potential at the solar surface, i.e. $\phi_{int}(R_{\odot},u)$=$\phi_{out}(R_{\odot},u)$, as follow:
\begin{equation}\label{eq:J2n}
  J_{2n}=\frac{R_{\odot}}{GM_{\odot}}\phi_{12n}\left(R_{\odot}\right)
\end{equation}

Applying this perturbation technique to stellar structure equations, \citet{mecheri2004} derived a convenient form of the Poisson equation for a general $n$ which is given as follow :
\begin{eqnarray}\label{eq:difphiint}
  \frac{d^{2}\phi_{12n}}{dr^{2}}+\frac{2}{r}\frac{d\phi_{12n}}{dr}-
  \left(2n\left(2n+1\right)+UV\right)\frac{\phi_{12n}}{r^{2}}=
  \nonumber\\
  U\left(\left(V+2\right)A_{2n}+r\frac{dA_{2n}}{dr}+B_{2n}\right)
\end{eqnarray}
which was obtained by combining linearized equations governing the equilibrium of rotating star in which only first order terms have been retained \citep{goldreich1968,ulrich1981a,ulrich1981b}. The quantities $U=4\pi\rho_{0}r^{3}/M_{r}$ and $V$=dln$\rho_{0}$/dln$r$, which refer to a spherical non-rotating Sun, are obtained from solar models through the density $\rho_{0}$ and the mass $M_{r}$ contained in a sphere of radius $r$ inside the Sun. For a solar angular velocity $\Omega(r,u)$, the quantities $A_{2n}$ and $B_{2n}$ are given by:
\begin{eqnarray}\label{eq:a2nb2n}
  &A_{2n}(r)&=\int_{-1}^{1}a_{2n}(u)\Omega(r,u)^{2}\texttt{d}u
  \nonumber\\&&=-\frac{1}{2n!}\frac{4n+1}{2^{2n+1}}\int_{-1}^{1}
  u\Omega(r,u)^{2}\frac{\texttt{d}^{2n-1}}{\texttt{d}u^{2n-1}}
  \left(u^{2}-1\right)^{2n}\texttt{d}u
  \nonumber\\
  &B_{2n}(r)&=\int_{-1}^{1}b_{2n}(u)\Omega(r,u)^{2}\texttt{d}u
  \nonumber\\&&=\frac{4n+1}{2}\int_{-1}^{1}
  \left(1-u^{2}\right)P_{2n}(u)\Omega(r,u)^{2}\texttt{d}u
\end{eqnarray}

Following closely the treatment of \citet{pijpers1998} using the Green's functions method, it is possible to derive from the above general differential equation~(\ref{eq:difphiint}), a general integral equation giving $\phi_{12n}$ at the surface of the Sun:
\begin{eqnarray}\label{eq:intphiint}
  &\phi_{12n}(R_{\odot})&=-\frac{R_{\odot}^{-2n}}
  {GM_{\odot}}\left[\frac{r^{2n}}{(2n+1)\psi_{2n}+r\psi_{2n}^{'}}\right]_{r=R_{\odot}}\times
  \nonumber\\&&
  \int^{R_{\odot}}_{0}r^{2}U\left(\left(V+2\right)A_{2n}+r\frac{dA_{2n}}{dr}+
  B_{2n}\right)\psi_{2n}\texttt{d}r
\end{eqnarray}
where $\psi_{2n}(r)$ is a regular solution at the origin (i.e. $\psi_{2n}(r)\propto r^{2n}$ as $r \rightarrow 0$) of equation~(\ref{eq:difphiint}) with a right hand side identical to zero and  $\psi_{2n}^{'}(r)$ is its derivative with respect to $r$. Finally, using equation~(\ref{eq:J2n}) and dimensionless variables $x=r/R_{\odot}$, $\omega^{2}=\Omega^{2}(R_{\odot}^{3}/GM_{\odot})$, $J_{2n}$ is given by:
\begin{eqnarray}\label{eq:intJ2n}
  &J_{2n}&=-\left[\frac{x^{2n}}
  {(2n+1)\psi_{2n}+x\psi_{2n}^{'}}\right]_{x=1}\times
  \nonumber\\
  &&\int^{1}_{0}\left(\left(x^{2}(U-4)U\psi_{2n}-x^{3}U\psi_{2n}^{'}\right)A_{2n}+
  x^{2}U\psi_{2n} B_{2n}\right)\texttt{d}x
  \nonumber\\
  &&=\int^{1}_{0}\int^{1}_{-1}F_{2n}(x,u)\omega(x,u)^{2}\texttt{d}u\texttt{d}x
\end{eqnarray}

The normalized integration kernel $F_{2n}(x,u)$ is therefore given by:
\begin{eqnarray}\label{eq:intF2n}
  &F_{2n}(x,u)&=-\left[\frac{x^{2n}}
  {(2n+1)\psi_{2n}+x\psi_{2n}^{'}}\right]_{x=1}\times
  \nonumber\\&&
  \left(\left(x^{2}(U-4)U\psi-x^{3}U\psi_{2n}^{'}\right)a_{2n}+x^{2}U\psi_{2n} b_{2n}\right)
\end{eqnarray}
Note that for $n=1$, equation~(\ref{eq:intJ2n}) reduces to equation~(23) of \citet{pijpers1998} in the case of general angular rotation $\omega(x,u)$ and to equation~(12) of \citet{gough1981} for a radially dependent angular rotation $\omega(x)$.

\section{Results and discussion}
\label{sec:results}
\begin{table*}
	\centering
	\caption{Values of solar gravitational moments $J_{2n}$ ($n$=1,2,3,4 and 5) computed using solar models from CESAM and ASTEC stellar evolution codes and rotation rates obtained from HMI and MDI helioseismic data, together with values from other authors also computed using helioseismic estimates of internal rotation.}
	\label{tab:tabJ2n}
    \begin{tabular}{lllllll}
		\hline
		 Authors & Rotation data & $J_{2}(\times10^{-7})$ & $J_{4}(\times10^{-9})$ &
        $J_{6}(\times10^{-10})$ & $J_{8}(\times10^{-11})$ & $J_{10}(\times10^{-12})$ \\
		\hline
Present work & SDO/HMI (CESAM) & ~2.211 & -4.252 & -1.282 & ~5.897 & -4.372 \\
             & SDO/HMI (ASTEC) & ~2.216 & -4.256 & -1.283 & ~5.901 & -4.375 \\
             & SoHO/MDI (CESAM) & ~2.204 & -4.064 & -1.136 & ~5.404 & -3.993 \\
             & SoHO/MDI (ASTEC) & ~2.208 & -4.069 & -1.137 & ~5.408 & -3.996 \\
             \hline
	  \cite{antia2008} & GONG & ~2.22 & -3.97 & -0.8 & ~1.1 & ~7.4 \\
                       & SoHO/MDI & ~2.18 & -4.70 & -2.4 & -0.8 & ~7.1 \\
	  \cite{mecheri2004} & SoHO/MDI & ~2.205 & -4.455 &  &  & \\
      \cite{roxburgh2001} & SoHO/MDI (ISM)& ~2.208 & -4.46 & -2.80 & ~1.49 & \\
                          & SoHO/MDI (CSM)& ~2.206 & -4.44 & -2.79 & ~1.48 & \\
      \cite{antia2000} & GONG+SoHO/MDI & ~2.18 & -4.64 &  &  & \\
      \cite{armstrong1999} & SoHO/MDI & ~2.22 & -3.84 &  &  & \\
      \cite{godier1999} & SoHO/MDI & ~1.6 &  &  &  & \\
      \cite{pijpers1998} & GONG+SoHO/MDI & ~2.18 &  &  &  & \\
      \cite{paterno1996} & IRIS+BISON+LOWL & ~2.22 &  &  &  & \\
      \cite{brown1989} & SPO/Fourier-Tachometer & ~1.7 &  &  &  & \\
      \cite{duvall1984} & KPNO/McMath-telescope & ~1.7 &  &  &  & \\
     \hline
	\end{tabular}
\end{table*}

The calculated values of $J_{2n}$ for $n$=1,2,3,4 and 5 together with previously published results also obtained using a helioseismic estimates of internal rotation are given in Table~\ref{tab:tabJ2n}, where a difference in sign convention has been taken into account concerning the results of \citet{armstrong1999} and \citet{antia2000}. They have been computed using equation~(\ref{eq:intJ2n}), in which the function $\psi_{2n}$ and the kernel $F_{2n}$ are evaluated using the quantities $U$ and $V$ from two solar models obtained from CESAM \citep{morel2008} and ASTEC \citep{dalsgaard2008} stellar evolution codes. For $\omega$, we use time-averaged two-dimensional rotation rates obtained from SDO/HMI helioseismic data of full-disk (fd\_V) dopplergrams available in the SDO HMI-AIA Joint Science Operations Center (JSOC) database covering the period between April 2010 and July 2020. For comparison purpose, we also compute $J_{2n}$ using rotation rates provided by the Michelson Doppler Imager (MDI) onboard of the Solar and Heliospheric Observatory (SoHO), available in the same database for the period between May 1996 and March 2008. This comparison is all the more interesting as, unlike previous contributions of Table~\ref{tab:tabJ2n}, it uses rotation rates obtained form an improved recent analysis of fd\_V MDI helioseismic data \citep{larson2015,larson2018} which corrects for several geometric effects during spherical harmonic decomposition as well as some other physical effects such as the distortion of eigenfunctions by the differential rotation and the horizontal displacement at the solar surface. The HMI fd\_V data, which requires less geometric corrections, have been processed exactly in the same manner as the MDI fd\_V data. The rotation rates for both datasets, have been calculated using two-dimensional regularized least-squares (RLS) inversions \citep{schou1998} of odd rotational splitting coefficients of $f$-modes and $p$-modes frequencies. Fig.~\ref{fig:figrot} shows superimposed time-averaged radial profiles at different latitudes of HMI (solid lines) and MDI (dashed lines) rotation. The two rotation profiles are very similar with only small differences at high latitude in the convective zone. However, a more pronounced difference can be noticed in deeper region inside the Sun below approximately $0.4R_{\odot}$. It should be noted that these two locations are regions in the Sun where rotation estimates are considered unreliable, but nevertheless we use them in our calculations in the absence of other alternatives. Table~\ref{tab:tabJ2n} shows that, for the same solar model, the calculated values of $J_{2n}$ from HMI and MDI rotation data have the same order of magnitude with however a slightly larger absolute values for HMI results. The difference is approximately of the order of 0.3\% for $J_{2}$ and increases for higher multipole moments to 4\% for $J_{4}$, 11\% for $J_{6}$, 8\% for $J_{8}$ and 9\% for $J_{10}$, presumably due to the difference in the rotation deep inside the Sun for $J_{2}$ and in the outer layers for higher multipole moments. Indeed, as already emphasized by \citet{antia2008}, high order multipole moments are predominantly determined from the contributions of the outer layers of the Sun where their integration kernels are principally concentrated as shown in Fig.~\ref{fig:figF2n} and \ref{fig:figF2nrt}  (for $n$=2,3,4 and 5), exhibiting substantial variation with latitude, with local minima and maxima positioned approximately at radial distances between $0.8R_{\odot}$ and $0.9R_{\odot}$. On the other hand, the major contribution to $J_{2}$ comes from deeper regions where the corresponding integration kernel (see Fig.~\ref{fig:figF2n} and \ref{fig:figF2nrt}, for $n$=1) exhibits its greatest value also at $r\approx0.77R_{\odot}$ principally at low latitudes around $34^{\circ}$.
Note that the sensitivity of high order multipole moments to the differential rotation in the outer layers of the Sun has been evidenced for $J_{4}$ by \citet{mecheri2004}, particularly the effect due to the presence of a subsurface radial gradient. More pronounced differences in the values of $J_{2n}$ have been found by \citet{antia2008} using GONG and MDI rotation rates (Table~\ref{tab:tabJ2n}) which, according to the authors, are the direct consequence of the differences between the measured splitting coefficients. For $J_{2}$, our result are in close agreement with most of the evaluations reported in Table~\ref{tab:tabJ2n}, except for those of \citet{godier1999,brown1989,duvall1984} which are considerably smaller. For \citet{duvall1984} and \citet{brown1989}, this difference is principally due to the very early helioseismic data used in the inference of internal rotation, restricted to regions close to the equator for the former. Surprisingly, \citeauthor{godier1999}'s value of $J_{2}$ is also largely inferior to the ones obtained by \citet{mecheri2004} and \citet{roxburgh2001} despite of using exactly the same rotation law. Higher order multipole moments $J_{6}$, $J_{8}$ and $J_{10}$ have the same order of magnitude as those of \citet{roxburgh2001} and \citet{antia2008}, with however sensitively different exact values. It is worth mentioning that \citeauthor{roxburgh2001}'s results have been obtained using a rotation model in a parametric form which roughly approximate the internal rotation inferred from helioseismology. Note from Table~\ref{tab:tabJ2n}, that for the same rotation data, our results from the two solar models are in very good agreement with insignificant differences inferior to $0.2\%$. Similar compatibility was found by \citet{roxburgh2001} for $J_{2}$, $J_{4}$, $J_{6}$ and $J_{8}$ computed using inverted (ISM) and calculated (CSM) solar models (see Table~\ref{tab:tabJ2n}). This compatibility is also verified when comparing the values of $J_{2}$ and $J_{4}$ obtained respectively by \citet{roxburgh2001} and \citet{mecheri2004} using distinct solar models but the same model of rotation of \citet{kosovichev1996}. Both authors pointed out that the differential rotation in the convective zone introduces only a diminution of $0.5\%$ of the value of $J_{2}$ with comparison to the one obtained for a Sun rotating uniformly at the rotation rate of the radiative interior. This indicates that the quadrupole moment $J_{2}$ is basically determined by a spherically averaged rotation whose departure from interior rotation is relatively small \citep{roxburgh2001}.

On he other hand, the sensitivity of high order multipole moments to the differential rotation in the convective zone makes them responsive to the observed temporal variation of the latitudinal component of the angular rotation \citep{howe2009} exhibiting changes either correlated or anti-correlated with magnetic activity \citep{antia2008}, whereas by contrast, $J_{2}$, which is more sensitive to the radiative zone rotation, do not present significant variation basically because the angular rotation in deeper layers inside the Sun do not show reliable temporal fluctuations. However, observational temporal changes of $J_{2}$ have been recently evidenced by \citet{rozelot2020} from the analysis of the perihelion precession measurements of several planets taken at different periods. \citeauthor{rozelot2020} reported a mean weighted value of $J_{2}=(2.17\pm0.06)\times10^{-7}$ which is very compatible with our results. We mention also the good compatibility of our results with the value $J_{2}=(2.25\pm0.09)\times10^{-7}$ deduced from the measurements of the precession of Mercury's perihelion obtained from ranging data of the MESSENGER (MErcury Surface, Space ENvironment, GEochemistry, and Ranging) spacecraft \citep{park2017}. They are however not compatible with the earlier values of $J_{2}=(1.8\pm5.1)\times10^{-7}$ and $J_{4}=(9.8\pm4.6)\times10^{-7}$ found by \citet{lydon1996} from the SDS (Solar Disk Sextant) balloon-borne experiment.

The calculated quadrupole moment $J_{2}$ gives an approximate estimate of the theoretical solar oblateness $\Delta_{\odot}$ via the formula $\Delta_{\odot}\approx(3/2)J_{2}+(\delta r/R_{\odot})$, where $\delta r/R_{\odot}=8.1\times10^{-6}$ \citep{dicke1970}, yielding $\Delta_{\odot}\approx8.43\times10^{-6}$. This value is in fair agreement with most of the observational oblateness estimates from the analysis of space-based solar limb shape measurements, namely by SoHO/MDI \citep{emilio2007}, SODISM (Solar Diameter Imager and surface Mapper) onboard of PICARD spacecraft \citep{irbah2014,meftah2015} and SDO/HMI \citep{meftah2016,irbah2019}. It is worth to note also its excellent agreement with the most accurate oblateness measurement to date $(8.35\pm0.15)\times10^{-6}$ obtained from RHESSI/SAS limb data \citep{fivian2008}.

Finally, the calculation of $J_{2n}$ and resulting $\Delta_{\odot}$ for all MDI and HMI rotation data available for an entire period of two solar cycles, can make possible to explore their temporal variation and possible relation to magnetic activity and therefore allow for a direct comparison with optical limb shape inference of solar oblateness. The study of the dynamic evolution of these quantities from model calculations is an ongoing work which will be the subject of a future publication.

\section{Conclusions}
\label{sec:conclusions}

The precise theoretical estimate of solar gravitational moment $J_{2n}$ is very important in many astrophysical applications. In this work, we have used new HMI solar rotation rates to calculate updated values of $J_{2n}$ (for $n$=1,2,3,4 and 5) by mean of a general integral equation derived in the framework of the theory of slowly rotating stars. The results revealed a good agreement with most of the earlier helioseismic estimates particularly for $J_{2}$ and $J_{4}$, whereas $J_{6}$, $J_{8}$ and $J_{10}$ agree as an order of magnitude but however differ in their exact values. On the other hand, the comparison with the calculation results obtained using MDI rotation rates yielded a difference of the order of $\approx0.3\%$ for the quadrupole moment $J_{2}$. This difference increases by one order of magnitude for higher order multipole moments indicating their greater sensitivity, as compared to $J_{2}$, to the differences between HMI and MDI rotation rates, particularly in the outer layers of the Sun. The calculated value of $J_{2}\approx2.21\times10^{-7}$ is in agreement with the observational value $J_{2}=2.25\times10^{-7}$ provided by the high precision measurements of the precession of Mercury's perihelion obtained from ranging data of the MESSENGER spacecraft. The resulting theoretical value of the solar oblateness $\Delta_{\odot}$ was found to be approximately equal to $8.43\times10^{-6}$ which is in perfect accordance with the most accurate space-based observational estimate of $8.35\times10^{-6}$ obtained by RHESSI/SAS. The dynamic evolution of $J_{2n}$ and $\Delta_{\odot}$ and its eventual correlation with magnetic activity during solar cycles 23 and 24 is an ongoing work for a planned subsequent contribution.

\section*{Acknowledgements}

This work has been performed with support of the Centre National d'Etudes Spatiales (CNES), Stanford University and the University of Hawai'i System. The HMI data used are courtesy of NASA/SDO and the HMI science teams. This work also utilises data from the SoHO/MDI instrument. SoHO is a project of international cooperation between ESA and NASA. The authors thank the referee for the constructive remarks and suggestions which helped improving the quality of the paper.

\section*{Data Availability}

All MDI and HMI rotation data used in this study are available online from the global helioseismology pipeline on the website of the Joint Science Operations Center (JSOC) at http://jsoc.stanford.edu/MDI/Global\_products.html for MDI and likewise at http://jsoc.stanford.edu/HMI/Global\_products.html for HMI, as cited in \citet{larson2015,larson2018}.

\bibliographystyle{mnras}
\bibliography{Mecheri-MNRAS-2021}
\appendix

\bsp	
\label{lastpage}
\end{document}